# Assignment of multiband luminescence due to the gallium vacancy–oxygen defect complex in GaN

Zijuan Xie,[1,2] Yu Sui,*[1] John Buckeridge,†[3] Alexey A. Sokol,‡[3] Thomas W. Keal,[4] and Aron Walsh§[,2,5]

[1]*Department of Physics, Harbin Institute of Technology, Harbin 15000, China*

[2]*Department of Materials, Imperial College London, London SW7 2AZ, United Kingdom*

[3]*Kathleen Lonsdale Materials Chemistry, Department of Chemistry, University College London, London WC1H 0AJ, United Kingdom*

[4]*Scientific Computing Department, Daresbury Laboratory, STFC, Daresbury, Warrington WA4 4AD, United Kingdom*

[5]*Department of Materials Science and Engineering, Yonsei University, Seoul 03722, Korea*

*\*suiyu@hit.edu.cn, †uccajab@ucl.ac.uk, ‡uccaaa0@ucl.ac.uk, §a.walsh@imperial.ac.uk*

*Oxygen is the most common unintentional impurity found in GaN. We study the interaction between substitutional oxygen ($O_N$) and the gallium vacancy ($V_{Ga}$) to form a point defect complex in GaN. The formation energy of the gallium vacancy is largely reduced in n-type GaN by complexing with oxygen, while thermodynamic and optical transition levels remain within the band gap. We study the spectroscopy of this complex using a hybrid quantum-mechanical molecular-mechanical (QM/MM) embedded-crystal approach. We reveal how a single defect center can be responsible for multiband luminescence, including the ubiquitous yellow luminescence signature observed in n-type GaN, owing to the coexistence of diffuse (extended) and compact (localized) holes.*

Gallium nitride (GaN) is an important semiconductor for optoelectronic and high-power devices such as light-emitting diodes and laser diodes.[1] In these devices, n-type doping of GaN is often achieved by incorporating oxygen (O) impurities that substitute a nitrogen (N) ion and donate free electrons to the conduction band.[2] Oxygen also tends to be present in undoped GaN, which is natively n-type.[3] In most n-type GaN samples, a yellow luminescence (YL) feature that has a peak at about 2.1–2.3 eV has been widely observed regardless of the growth technique used, which varies from molecular beam epitaxy (MBE)[4], and metal-organic chemical vapor deposition (MOCVD) [5,6,7] to hydride vapor phase epitaxy (HVPE)[7,8,9]. In some samples, the YL is accompanied by a red luminescence (RL)[7,10] or a green luminescence (GL)[8,11].

The source of YL is the subject of an on-going debate in the literature. Ogino and Aoki first proposed that YL is due to a radiative transition from a shallow donor with a depth of 25 meV to a deep acceptor with a depth of 860 meV.[12] From first-principles density-functional theory (DFT) calculations, Neugebauer and Van de Walle attributed this deep acceptor to a gallium vacancy ($V_{Ga}$, or related complex), which has a transition level 1.1 eV above the valence band maximum (VBM), while carbon (C) substituting for N ($C_N$) is a shallow acceptor.[13] Later calculations using hybrid DFT, predicted $C_N$ to be a deep acceptor which has an ionization energy of 0.90 eV and is another plausible source of YL.[14] However, by combining hybrid DFT and experimental measurements, Demchenko *et al.* found that the $C_N$-$O_N$ complex is energetically favorable and its (0/+) thermodynamic transition deep-donor level is at 0.75 eV above the VBM, which gives rise to the YL.[15] They also ruled out $C_N$ and $V_{Ga}$-$O_N$ as the sources of YL. Lyons *et al.*[16], again using hybrid DFT, predicted a 2.27 eV emission peak for radiative hole capture by $V_{Ga}^{3-}$, which, however, has a high formation energy. Instead, they found that donor impurities significantly lower $V_{Ga}$ transition levels, and that $V_{Ga}$-3H and $V_{Ga}$-$O_N$-2H complexes may give rise to YL. Experimentally, YL has been attributed to $V_{Ga}$ (and/or related complexes)[5,9,8] or C-related defects[11,17], or both[6], or neither[7]. There is experimental evidence that localization of holes is involved in YL.[6,18,9] Xu *et al.* found that a GaN film with a shorter positron diffusion length exhibits stronger YL, and suggested that the extremely strong space localization effect of holes is the vital factor to enhance the YL efficiency.[6] Reshchikov attributed YL and GL bands in high-purity undoped GaN to an acceptor binding one and two holes, respectively.[9] $V_{Ga}$ complexes have also been studied as non-radiative recombination centers by calculations[19] and experiments[20,21].

The sources of the RL and GL bands that are often observed together with YL are also unclear. Some PL spectra show multiple YL and GL bands.[22] They are often suggested to be caused by the same defect.[23] From time-resolved photoluminescence (PL) spectra, Reshchikov *et al.* observed a GL band disappear at long-time decays and YL and RL bands become dominant, while in the steady-state PL, the GL band is not observed in the time-resolved PL spectra but the YL and RL bands can be resolved.[24] Bozdog *et al.* observed overlapping RL, YL and GL bands in GaN grown by HVPE. They followed a two-stage model, in which the luminescence arises from hole capture at the defect after an electron capture process.[18] Dı́az-Guerra *et al.* found two YL bands peaked at 2.22 and 2.03 eV, and a RL band at about 1.85 eV that dominates spectra recorded for long delay times.[10] Reshchikov *et al.* first observed the fine structure of the RL band with a maximum at 1.8 eV and a zero-phonon line (ZPL) at 2.36 eV.[25] Then they attributed this fine structure to a YL band often covered by the RL band.[24]

$V_{Ga}$ as a native defect has been widely discussed, based on DFT calculations, as the reason for YL, as it is an acceptor and thus has lower formation energy in n-GaN.[13] The calculated formation energies of neutral $V_{Ga}$ under N-rich condition by different functionals are close, about 6.5 eV[26,27,28,29,30,31]. However, the transition levels of $V_{Ga}$ across the band gap differ, which render different formation energy of $V_{Ga}$ near the CBM, when in a -3 charge state. Transition levels of $V_{Ga}$ are closer to the conduction band when comparing results from local (or semi-local) DFT functionals to hybrid functionals that are capable of describing hole localization.[16] The associated formation energy of $V_{Ga}$ at the CBM vary from about 0 eV from (local) LDA[26] to 4.4 eV from (hybrid) HSE[16]. The higher formation energy from hybrid-DFT calculations indicates that $V_{Ga}$ may not be thermodynamically abundant even under n-type GaN. However, the formation of $V_{Ga}$ is more likely if it forms a complex with impurities like O or H, as has been shown in calculations.[13,16,32] Indeed, positron annihilation spectroscopy has been used to verify the decoration of Ga vacancies in GaN by O or H.[33,34]

Until now the origin of YL is not clear. Which transition in which defect or complex accounts for YL is in constant debate. Previous computational or experimental studies are inconclusive, and sometimes contradictory. $O_N$ is common in n-GaN as a dopant or impurity, and the YL in n-GaN has been shown to be related to $V_{Ga}$, which is likely to form a complex with a donor impurity such as O. In this Letter, we report a quantum mechanical/molecular mechanical (QM/MM) embedded cluster study of $V_{Ga}$-$O_N$ to illustrate its optical and electronic properties, especially its effect on luminescence.

The QM/MM embedded cluster technique employed describes accurately localized states in ionic solids, where charged and strongly dipolar species predominantly interact via long-range electrostatic and short-range exchange forces[35,36]. This method and a plethora of closely related embedding approaches[37] split extended systems into an inner region containing the central defect, described using molecular QM theories, and its surroundings, which are only slightly perturbed by the defect, modelled with MM approaches.

Our choice of QM methodology is DFT; the MM simulations employ polarizable shell model interatomic potentials; and the interface between these two regions is based on cation-centred semi-local pseudopotentials. The inner cluster of 116 atoms of GaN centred on the defect is treated using (i) the second-generation thermochemical hybrid exchange and correlation (XC) density functional B97-2,[38] which is similar to those commonly used in recent plane-wave supercell calculations (21% exact exchange compared with 25% for PBE0[39] or HSE06[40]), (ii) the SBKJC small-core pseudopotentials on Ga[41] within the cluster and large-core refitted pseudopotentials[42,43] in the interface that provide a short-range contribution to the embedding potential on the defect, and (iii) the atomic basis set of def2-TZVP quality on N[44] and matching SBKJC basis on Ga[41]. For comparison, we use a second hybrid XC density functional employing 42% exact exchange (BB1k),[45] fitted to reproduce kinetic barriers and thermochemical data, which gives a more accurate description of electron localization than B97-2[43,46]. The QM region of radius 6.8 Å is embedded in an outer cluster of radius 30 Å, which is treated with an MM level of theory using two-body interatomic potentials parameterized to reproduce GaN bulk structure and physical properties[43,47]. The method has been implemented in the ChemShell package[35,36,48] that employs Gamess-UK[49] for the QM and GULP[50] for MM single point energy and gradient calculations. Further technical details are discussed elsewhere[35,51]. This method has been applied successfully to treat defects in AgCl[52], ZnO[53] and GaN[43] and the band alignment of polymorphs of $TiO_2$[43,54].

The formation energy of a point defect $X(E_D[X])$ is determined from the grand-canonical expression:

$$E_D[X] = \Delta E(X) - \sum_i n_i \mu_i + qE_F, \tag{1}$$

where $\Delta E(X)$ is the difference in energy between the embedded cluster with and without $X$, $n_i$ is the number of atoms of species $i$ added ($n_i > 0$) or subtracted ($n_i < 0$) in forming $X$, $\mu_i$ is the chemical potential of species $i$, $q$ is the charge of $X$, and $E_F$ is the Fermi energy. $\mu_i$ depends on the experimental growth condition, which can be either N or Ga rich. The binding energy of $V_{Ga}$-$O_N$ complex is determined as[55]:

$$E_D[V_{Ga}^m] + E_D[O_N^n] \rightarrow E_D[(V_{Ga} - O_N)^p] + (m + n - p)E_F + E_B, \tag{2}$$

where $m$, $n$, and $p$ are charge states and $E_B$ is the binding energy of the complex; a positive value means the complex is likely to form.

GaN adopts a wurtzite structure ($C_{3v}$ point group), where each Ga (N) is surrounded by four N (Ga), one in the axial direction, the other three equivalently in a basal plane. We thus consider two geometries of the $V_{Ga}$-$O_N$ complex: $V_{Ga}$ with a O substituting an axial N or a basal N. The formation energies of these two geometries are almost identical, as shown in Fig. 1. The formation energy difference between Ga-rich and N-rich conditions comes from the chemical potential of O, taken from $Ga_2O_3$ and $O_2$, respectively. The binding energies of the $V_{Ga}$-$O_N$ complex when calculated using either functional are higher than 2 eV at the CBM. We have shown in a previous study[56] that $O_N$ will form spontaneously for any value of Fermi energy across the gap and, indeed, in the conduction band, in sharp contrast to $V_{Ga}$, (see Fig. 2) which have formation energies of at least 3 eV at the CBM. From the computed point defect formation energies, we therefore expect oxygen to be easily incorporated in the material, with practically no compensation by the formation of $V_{Ga}$. We note, however, that vacancies can be introduced, either by non-equilibrium processes or by diffusion from polar surfaces, and that the high binding energy then indicates that all available pairs of vacancies and oxygens will form complexes.

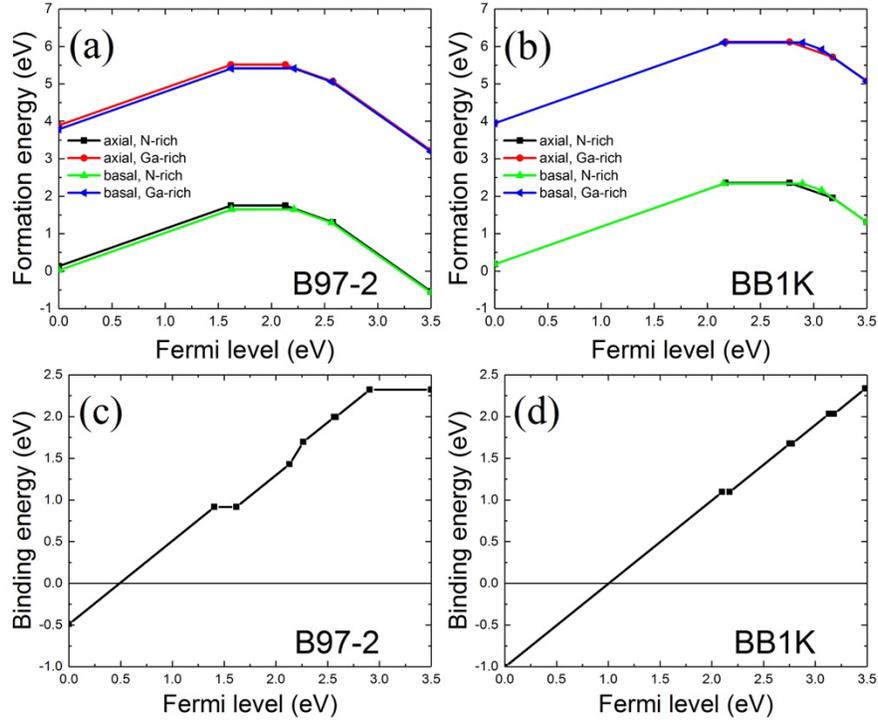

FIG. 1. Formation energies (a, b) and binding energies (c, d) of $V_{Ga}$-$O_N$ calculated using the B97-2 (a, c) and BB1K (b, d) XC functionals. Binding energies are calculated using the formation energies of axial $O_N$, under N-rich conditions; positive values represent favorable binding.

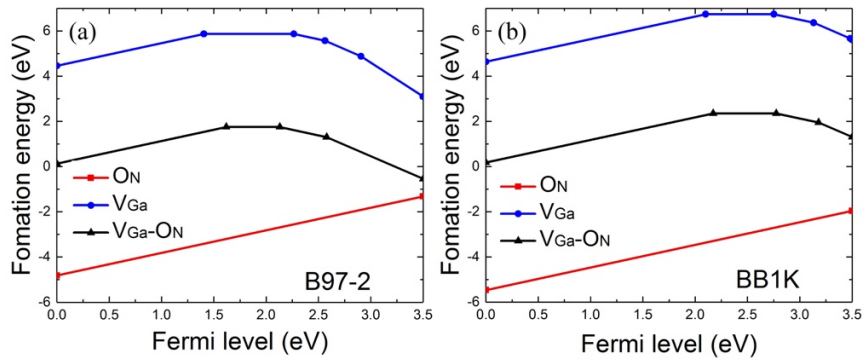

FIG. 2. Formation energies of isolated $V_{Ga}$, $O_N$ and axial $V_{Ga}$-$O_N$ defects calculated using the B97-2 (a) and BB1K (b) XC functionals.

Since O is close to N in size and chemistry, $V_{Ga}$-$O_N$ shares some properties with isolated $V_{Ga}$. Fig. 2[56] shows similar transition levels between $V_{Ga}$ and $V_{Ga}$-$O_N$, which was also found by Lyons *et al.*[16] We have shown in a previous study that the -2, -1, 0 and +1 states of $V_{Ga}$ can bind one, two, three, and four holes on nearest N; here we find $V_{Ga}$-$O_N$, too, can bind one, two and three holes on nearest N at -1, 0, +1 states, respectively.

Before discussing the optical properties of $V_{Ga}$-$O_N$, we point out that a defect can have two co-existing electronic states: a compact state and a diffuse state. Often, we only refer to the compact nature of a defect state and consider ionization to or from bands, as shown in Fig. 3. However, in real dielectrics, point defects can trap one or more charge carriers in shallow hydrogenic states (such as the diffuse hole state shown in Fig. 3), especially when the potential for forming such a diffuse state is an attractive Coulombic well. Such a carrier in a diffuse s-like state has an energy level shifted into the gap relative to the appropriate band edge (depending on the defect charge state), and the ground state of the defect can be either its diffuse state, for example $(V_{Ga} - O_N)^{-2} + h^+$ (diffuse), or its counterpart compact state, $(V_{Ga} - O_N)^{-1}$. Transitions can happen between these two states, and the state that will be observed is subject to both the property of the defect and the kinetics of the experiment.

For the $V_{Ga}$-$O_N$ complex we are studying here, we calculate radiative transitions of holes to negatively charged states (see Fig. 3). We thus consider the optical transition of holes from either the valence band, or a diffuse s-like hole level above the valence band, to the negative states of this complex. We have shown in a previous work[56], using effective mass theory, that a hole trapped by a singly charged state has an energy 0.20 eV lower than holes in the valence band, while a hole trapped by a doubly charged state has an energy 0.40 eV lower than the valence band.

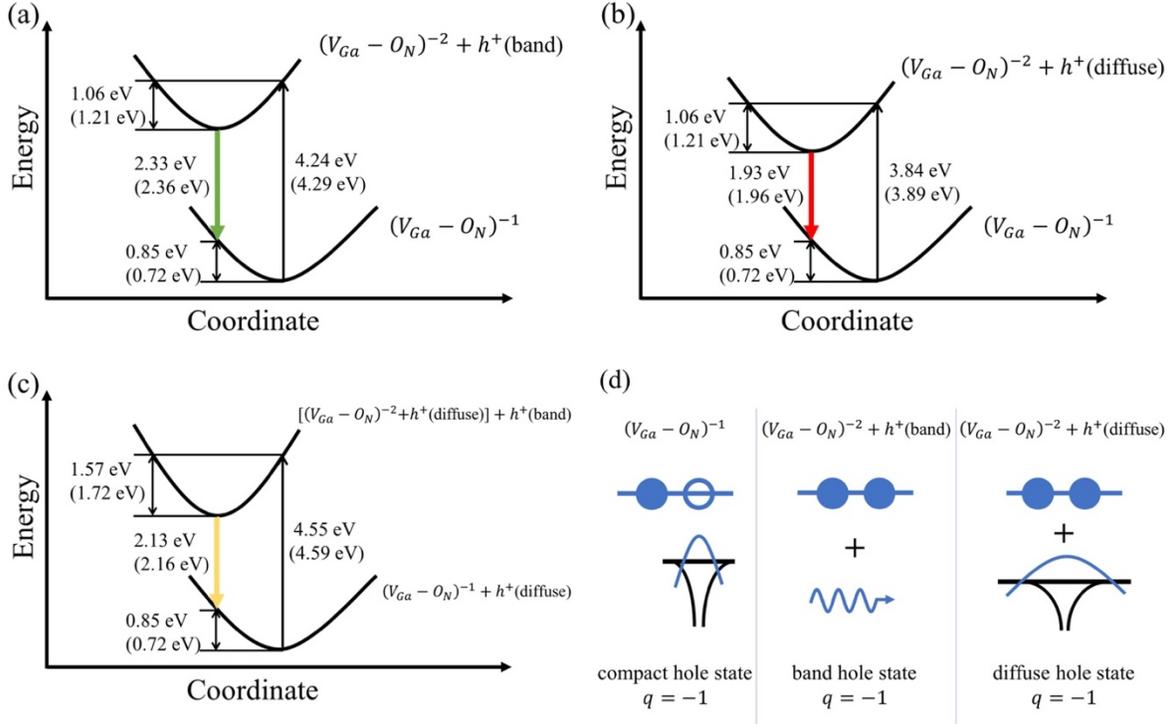

FIG. 3. Configuration-coordinate diagram describing the transition of a hole from the (a) valence band ($h^+$(band)) or (b) a corresponding diffuse hole ($h^+$(diffuse)) to the -2 state. The transition between two diffuse states is shown in (c). Note that the diffuse hole attracted by different charge states ($q$) has different energy levels, hence we have, $h^+$(diffuse, 2-) and $h^+$(diffuse, 1-). The values out of braces correspond to complexes involving axial O and in those in braces to complexes involving basal O. A comparison between a compact hole state (trapped in the local defect potential), a band hole state (free in the valence band), and a diffuse hole state (hydrogenic orbit bound by the charged defect) is shown schematically in (d).

For $V_{Ga}$-$O_N^{2-}$, the radiative hole capture from the VBM has a green emission peak at 2.33 eV for axial O and 2.36 eV for basal O. If the hole is captured from the diffuse state of $V_{Ga}$-$O_N^{2-}$, which is 0.40 eV above the VBM, the transition peak is at 1.93 eV (axial O, 1.96 eV for basal O), which is RL. When both the initial -2 state and the final -1 state of the hole transition are in their diffuse states, the corresponding transition peak is at 2.13 eV (axial O, 2.16 eV for basal O), corresponding to YL. For $V_{Ga}$-$O_N^{1-}$, the radiative hole capture from the VBM has a yellow emission peak at 2.10 eV for axial O and 2.20 eV for basal O. If the hole is captured from the diffuse state of $V_{Ga}$-$O_N^{1-}$, which is 0.20 eV above the VBM, the transition peak is at 1.90 eV (axial O, 2.00 eV for basal O).

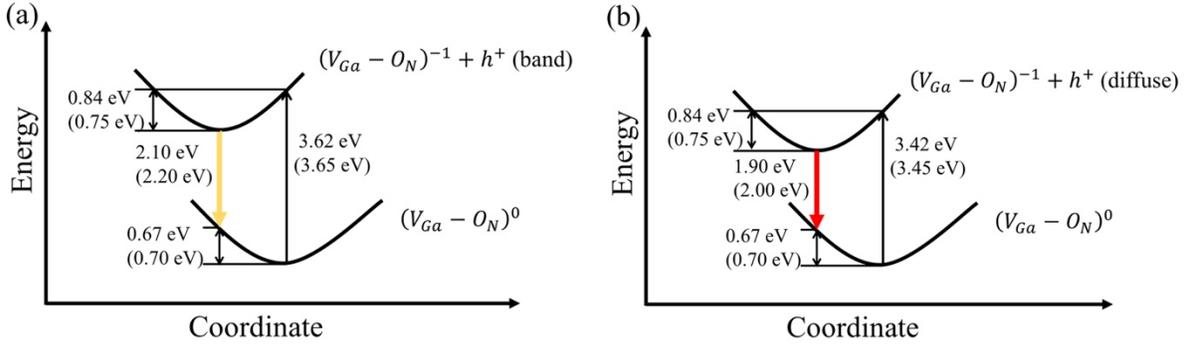

FIG. 4. Configuration-coordinate diagram for the transition of a hole from the valence band (a) or a corresponding diffuse level (b) to the -1 state. The values out of braces are from axial O and in them are from basal O.

The energy difference between GL and YL peaks from experiments in PL spectra is 0.26 eV[8], 0.2 eV[57] or 0.29 eV[22], which is comparable to our results. Sedhain identified two peaks at 2.13 and 2.30 eV and attributed them to two geometries of the $V_{Ga}$-$O_N$ complex: the axial configuration and basal configuration.[58] In our calculations, we find similar peaks but we do not find such a big difference between axial and basal configurations; however, the 0.20 eV change between compact and diffuse states is comparable. Reshchikov *et al.* observed a transformation of the YL into the GL band, and attributed the bands to two charge states of the same dominant acceptor,[8] which is consistent with our results where compact hole capture by -2 and -1 state of $V_{Ga}$-$O_N$ give rise to GL and YL, respectively. Later they also observed that the lifetime of the GL band stays unchanged up to 70 K and then increases by a factor of 20 at 300 K, and attributed the change to the existence of an excited state.[9] We find that this lifetime increase of GL is a result of trapped hydrogenic diffuse holes being released to the valence band as the temperature increases, which increases the probability of occurrence of the first process in Fig. 3. Concurrently, the lifetime of YL is decreased by a reduction in the rate of the last process in Fig. 3, while the occurrence of the first process in Fig. 4 is also reduced due to the increase in the Fermi level with temperature, as observed in their experiment.[9] In unintentionally doped GaN layers grown by MBE, Reshchikov *et al.* observed a RL band with a maximum at about 1.88 eV and a GL band with a maximum at about 2.37 eV.[20] Our calculated RL and GL peaks are very close (see Figs. 3 and 4). They speculate that the defects are partially nonradiative (with strong electron-phonon coupling) and related to Ga atoms. Indeed, gallium vacancy complexes have been studied as a cause of nonradiative recombination[19]; the large lattice distortion by hole capture has been reported previously[59] and is seen in our calculations.

Reshchikov *et al.* have resolved a YL band which has a maximum at 2.20 eV and zero phonon line (ZPL) at 2.57 eV, [7,21,24,60], or another YL band which has a maximum at 2.10 eV and ZPL at 2.36 eV.[24] The difference between band maximum and ZPL is the relaxation after the carrier capture process, which in the above YL bands are 0.37 eV and 0.26 eV. In our calculations, the relaxation energy ranges from 0.67 eV to 0.85 eV. The small relaxation resolved from the above experiments is quite unlikely according to both previous calculations[16,59] and our findings. The N surrounding the Ga vacancy relax strongly upon binding or releasing a hole when YL takes place. We suggest that a reinvestigation of the tails in the experimental PL spectra is called for.

In the literature, there are many discussions on electron traps that give rise to YL, by $C_N$[14], $C_N$-$O_N$[15] or $V_{Ga}$ complexes[16]. However, the transitions claimed to cause YL emission are all close to the valence band. For example, Lyons *et al.* identified electron capture by +1 state of $V_{Ga}$-3H and $V_{Ga}$-$O_N$-2H complexes to be the origin of YL, while their transition levels of (+/0) is less than 1 eV above the VBM where it is not n-type. Indeed, if electron capture is to give rise to an emission peak in the visible region, the defect level must be in the lower half of the band gap. Instead, our calculation shows that YL bands arise from hole capture following photoexcitation of electrons at transition levels near the CBM.

In conclusion, $V_{Ga}$-$O_N$ has a much lower formation energy compared to isolated $V_{Ga}$, while the transition levels of both defects in the band gap are quite similar. While $V_{Ga}$-$O_N$ and $V_{Ga}$ have similar hole capture peaks, $V_{Ga}$-$O_N$ is responsible for luminescence peaks observed in PL spectra due to its lower formation energy and thus higher concentration. The YL and accompanied RL and GL are explained as originating from transitions involving different charge states of $V_{Ga}$-$O_N$, taking into account processes involving bound diffuse holes and delocalised band holes. The model explains why YL, GL and RL bands are often observed in combination, and why multiple peaks are found for red and yellow luminescence.

Z. X. thanks the China Scholarship Council (CSC) for support. We are grateful to Alex Ganose and Stephen Shevlin for technical help, and Chris Van de Walle and Audrius Alkauskas for useful discussions. The EPSRC is acknowledged for funding (EP/K038419; EP/I03014X; EP/K016288). A. W. was supported by the Royal Society. Computational resources were provided through the Materials Chemistry Consortium on EPSRC grant number EP/L000202. We also acknowledge PRACE for awarding us access to the ARCHER supercomputer, UK.